\documentstyle[preprint,aps]{revtex}
\draft
\begin{document}
\title{
Composite Fermions and Landau Level Mixing in the Fractional Quantum
Hall Effect}
\author{V. Melik-Alaverdian and N.E. Bonesteel}
\address{
National High Magnetic Field Laboratory and Department of Physics,
Florida State University, Tallahassee, FL 32306-4005}
\maketitle
\begin{abstract}
The reduction of the energy gap due to Landau level mixing,
characterized by the dimensionless parameter $\lambda = (e^2/\epsilon
l_0)/\hbar\omega_c$, has been calculated by variational Monte Carlo
for the fractional quantum Hall effect at filling fractions $\nu=1/3$
and 1/5 using a modified version of Jain's composite fermion wave
functions.  These wave functions exploit the Landau level mixing
already present in composite fermion wave functions by introducing a
partial Landau level projection operator.  Results for the energy gaps
are consistent with experimental observations in $n$-type GaAs, but we
conclude that Landau level mixing alone cannot account for the
significantly smaller energy gaps observed in $p$-type systems.
\end{abstract}


The quantum Hall effect (QHE) occurs when a two dimensional electron
gas placed in a strong transverse magnetic field develops an energy
gap.  In the fractional QHE this energy gap is entirely due to
many-body correlations between the electrons \cite{qhe}.  It is
generally believed that the essential physics of the fractional QHE
can be understood in the limit where the cyclotron energy
$\hbar\omega_c = \hbar e B /m^*c$ is much greater than the Coulomb
energy scale $e^2/\epsilon l_0$ \cite{qhe}. Here $B$ is the applied
magnetic field, $m^*$ is the effective mass of the electron or hole,
$\epsilon$ is the dielectric constant and $l_0 = \sqrt{\hbar c/eB}$ is
the magnetic length.  In this limit the kinetic energy is completely
quenched, the electrons (or holes) are entirely in the lowest Landau
level provided the filling fraction is less than 1, and the Coulomb
energy is the only energy scale in the problem.

Jain has constructed a class of trial wave functions based on the idea
that the fractional QHE can be viewed as an effective integer QHE for
composite fermions --- electrons bound to an even number of
statistical flux quanta \cite{jain}.  Although Jain's wave functions
naturally explain many features of the observed fractional QHE
hierarchy, they suffer from a serious problem from the point of view
of computation.  With the exception of the `parent' ground states,
those with Landau level filling fractions $\nu = 1/q$ where $q$ is an
odd integer, Jain's wave functions are not entirely in the lowest
Landau level \cite{jain}.  Even for these parent filling fractions the
excited state wave functions constructed using composite fermions, in
particular the quasielectron wave function, also suffer from intrinsic
Landau level mixing \cite{trivedi,neb}.  Of course in real experiments
there will always be some Landau level mixing characterized by the
dimensionless parameter $\lambda = (e^2/\epsilon l_0)/\hbar\omega_c$.
The intrinsic Landau level mixing of Jain's wave functions is
therefore not entirely unphysical.  The problem is that the amount of
Landau level mixing in a given composite fermion wave function is
fixed, whereas in real experiments it depends on the parameter
$\lambda$.

Recently Manoharan {\it et al.} \cite{manoharan} have measured the
fractional QHE energy gaps in high quality two dimensional $p$-type
systems realized in GaAs/Al$_x$Ga$_{1-x}$As quantum wells.  At filling
fraction $\nu=1/3$ the energy gap they observe is $\Delta_h \simeq
0.023 e^2/\epsilon l_0$, roughly a factor of two smaller than the
corresponding energy gap $\Delta_e \simeq 0.05 e^2/\epsilon l_0$
observed in high quality $n$-type systems \cite{du,willett}. For
typical two dimensional carrier densities in GaAs heterostructures
$\lambda \simeq 1$ for $n$-type systems while $\lambda \simeq 5$ for
$p$-type systems.  This led Manoharan {\it et al.} to suggest that the
reduced energy gap they observed in $p$-type systems might be due to
Landau level mixing.

Motivated by these experiments we have calculated the $\lambda$
dependence of the energy gap for $\nu=1/3$ and 1/5 using a new class
of variational wave functions.  These wave functions are constructed
by applying a partial Landau level projection operator to Jain's
composite fermion wave functions.  In this way we have exploited the
intrinsic Landau level mixing already present in these wave functions
by introducing a variational parameter which controls the amount of
this mixing.  Previously Yoshioka \cite{yoshioka} studied the effect
of Landau level mixing on the excitation spectra of the fractional QHE
using exact diagonalization on small systems, including both the
lowest and first excited Landau levels, for $\nu=1/3$. Where it is
possible to compare, our variational results are in good agreement
with Yoshioka's results; however, since our calculations are based on
variational Monte Carlo we can study significantly larger systems, as
well as filling fraction $\nu=1/5$.  The results we have obtained for
the energy gap are consistent with experiments in $n$-type GaAs, but
we conclude that Landau level mixing alone cannot account for the
smaller energy gaps observed in $p$-type systems.

We work in the spherical geometry introduced by Haldane
\cite{haldane}.  In this geometry the electrons, which are taken to be
fully spin polarized, are confined to the surface of a sphere of
radius $R$ and move in the magnetic field of a magnetic monopole
placed at the center of the sphere.  The magnetic field strength at
the surface of the sphere is $B = S (l_0/R)^2$ and the field is
described by the vector potential ${\bf A} = {\bf e}_\phi(\hbar S/eR)
\cot\theta$, where  $2S = q(N-1)$ for $\nu=1/q$.  In this gauge the
appropriate generalization of Laughlin's ground state wave function
is\cite{haldane}
\begin{eqnarray}
\psi \propto \prod_{i<j}(u_iv_j - v_iu_j)^q =
\prod_i u_i^{q(N-1)}\prod_{i<j}(z_i-z_j)^q.
\end{eqnarray}
Here $u_i = \cos\theta_i/2 \exp -i\phi_i/2$ and $v_i =
\sin\theta_i/2\exp i\phi_i/2$ are the spinor coordinates of the
$i^{th}$ electron, where $\theta_i$ and $\phi_i$ are the spherical
coordinates of the electron, $z_i = v_i/u_i$ is the complex
stereographic coordinate and $N$ is the total number of electrons.  We
have introduced the stereographic coordinates $z$ in order to write
$\psi$ in a way which is formally similar of Laughlin's droplet wave
function.  Throughout this paper all wave functions are understood to
be normalized and the proportionality sign will be used whenever
necessary. Jain showed that $\psi$ can also be written
\cite{jain}
\begin{eqnarray}
\psi &\propto& \prod_i u_i^{q(N-1)}\prod_{i<j}(z_i-z_j)^{q-1}\left|
\begin{array}{ccccc}
1 & z_1 & ... & z_1^{N-2} & z_1^{N-1} \\
\vdots & \vdots  &  &\vdots &\vdots  \\
1 & z_N & ... & z_N^{N-2} & z_N^{N-1} \\
\end{array}\right|.
\label{groundstate}
\end{eqnarray}
In this form the determinant corresponds to one filled {\it
pseudo}-Landau level of composite fermions, where the flux quanta
bound to the electrons are represented, roughly, by the
$\prod_{i<j}(z_i-z_j)^{q-1}$ Jastrow factor \cite{jain}.

To determine the energy gap we must calculate the energy difference
between $\psi$ and the appropriate excited state wave function.  The
excited state we have used is constructed by promoting a composite
fermion from the lowest pseudo-Landau level to the first excited
pseudo-Landau level.  It is possible to construct an entire low energy
band of excited states in this way \cite{dev}.  Here we are interested
in the state constructed by removing a composite fermion from the
lowest pseudo-Landau level at the bottom of the sphere and
reintroducing it into the first excited pseudo-Landau level at the top
of the sphere.  The resulting wave function is
\begin{eqnarray}
\psi^\prime &\propto& \prod_i u_i^{q(N-1)}\prod_{i<j}(z_i-z_j)^{q-1}\left|
\begin{array}{ccccc}
1 & z_1 & ... & z_1^{N-2} & \frac{\overline z_1}{1+|z_1|^2} \\
\vdots & \vdots  &  &\vdots &\vdots  \\
1 & z_N & ... & z_N^{N-2} & \frac{\overline z_N}{1+|z_N|^2}  \\
\end{array}\right|.
\label{unprojected}
\end{eqnarray}
This wave function describes a state with a charge $+e/q$ quasihole at
the bottom of the sphere, and a charge $-e/q$ quasielectron at the top
of the sphere.  The total angular momentum quantum number of
$\psi^\prime$ is $l= N$ and so it is orthogonal to the $l=0$ ground
state.  In the $N\rightarrow\infty$ limit $\psi^\prime$ describes a
state with a quasielectron--quasihole pair with infinite separation.
The energy gap for creating such a pair is precisely the energy gap
which appears in the activated temperature dependence of the
longitudinal resistance.

The energy difference between $\psi^\prime$ and $\psi$ has been
calculated previously for $\nu=1/3$ and was found to be $\Delta
\simeq 0.05 e^2/\epsilon l_0 + 0.16 \hbar\omega_c$
\cite{neb}.  The contribution to $\Delta$ which is proportional to
$\hbar\omega_c$ comes from the nonzero overlap of Jain's quasielectron
wave function with the first excited Landau level.  We have exploited
this intrinsic Landau level mixing in $\psi^\prime$ by introducing a
partial Landau level projection operator and considering the wave
function
\begin{equation}
\psi^\prime_\alpha \propto [1+(\alpha-1)P_{\rm lll}]\psi^\prime.
\label{ppo}
\end{equation}
Here $P_{lll}$ is an operator which projects fully into the lowest
Landau level.  When $\alpha = 1$ $\psi^\prime_\alpha$ is simply Jain's
unprojected wave function while in the limit $\alpha\rightarrow\infty$
$\psi^\prime_\alpha$ becomes the fully projected version of Jain's
wave function. The parameter $\alpha$ can therefore be used as a
variational parameter to control the intrinsic Landau level mixing
present in $\psi^\prime$.  For the simple case considered here (one
quasielectron) it is possible to perform the partial projection
analytically with the following result
\begin{eqnarray}
\psi^\prime_\alpha &\propto& \prod_i u_i^{q(N-1)}\prod_{i<j}(z_i-z_j)^{q-1}
\left|
\begin{array}{ccccc}
1 & z_1 & ... & z_1^{N-2} & \frac{\overline z_1}{1+|z_1|^2} +
\frac{(1-\alpha)(q-1)}{2S+2}
\sum_{i\ne 1}
\frac{1}{z_i-z_1} \\
\vdots & \vdots  &  &\vdots &\vdots  \\
1 & z_N & ... & z_N^{N-2} & \frac{\overline z_N}{1+|z_N|^2} +
\frac{(1-\alpha)(q-1)}{2S+2}
\sum_{i\ne N}
\frac{1}{z_i-z_N} \\
\end{array}\right|.
\end{eqnarray}
Both $\psi$ and $\psi^\prime_\alpha$ have the conventional form of a
Jastrow factor multiplying a Slater determinant.  We have applied
standard variational Monte Carlo techniques for such wave functions to
calculate their properties in what follows \cite{ceperely}.

Before proceeding it is necessary to discuss the finite thickness of
the two dimensional electron or hole gas.  The wave functions of
electrons or holes in the lowest subband of a two dimensional system
have a finite extent perpendicular to the plane of the system.  This
`thickness' has the effect of softening the short-range part of the
Coulomb interaction. Since it is precisely the short-range
interactions which determine the energy gap in the fractional QHE it
is crucial to include this effect in any realistic calculation before
comparing with experiment.  Following Yoshioka \cite{yoshioka} and
Zhang and Das Sarma \cite{zhang} the thickness corrections have been
included using the Fang-Howard wave function for the lowest subband in
an inversion layer,
\begin{equation}
\xi(w) = \left(\frac{b^3}{2}\right)^{1/2} w e^{bw/2},
\end{equation}
where $w$ is the coordinate perpendicular to the two dimensional gas.
The finite thickness then has the effect of modifying the Coulomb
interaction so that the effective electron---electron interaction is
\begin{equation}
U(r) = \frac{e^2}{\epsilon l_0}
\int_0^\infty dw^\prime \int_0^\infty dw |\xi(w)|^2
\frac{1}{\sqrt{r^2+(w-w^\prime)^2}} |\xi(w^\prime)|^2.
\label{mci}
\end{equation}
The thickness is characterized by the dimensionless parameter $\beta =
1/(bl_0)$.  For the systems we are considering here $\beta
\simeq 1$.

To calculate the dependence of $\Delta$ on $\lambda$ for a given value
of $\beta$ we must minimize the total energy of $\psi^\prime_\alpha$
as a function of $\alpha$.  This calculation is greatly simplified by
the following observation.  Because $\psi^\prime$ contains only a
single composite fermion in the first excited pseudo-Landau level it
can be decomposed as follows:
\begin{equation}
\psi^\prime = \gamma\psi^\prime_0 + (1-\gamma^2)^{1/2}
\psi^\prime_1.
\end{equation}
Here $\psi^\prime_0$ is the projected state with all $N$ electrons in
the lowest Landau level and $\psi^\prime_1$ is orthogonal to
$\psi^\prime_0$ with $N-1$ electrons in the lowest Landau level and 1
electron in the first excited Landau level.

It follows from (4) and (8) that the partially projected state can be
written
\begin{equation}
\psi^\prime_\alpha \propto \alpha\gamma\psi^\prime_0 +
(1-\gamma^2)^{1/2}\psi^\prime_1.
\label{decomp2}
\end{equation}
If we define the expectation value of the Coulomb interaction in units
of $e^2/\epsilon l_0$ to be $V[\alpha]
={\langle\psi_\alpha^\prime|V|\psi_\alpha^\prime\rangle}/(e^2/\epsilon
l_0)$, where $V =
\sum_{i<j} U(r_{ij})$ with $U(r)$ as defined in (\ref{mci}),
then from (\ref{decomp2}) it follows that
\begin{eqnarray}
V[\alpha] =
\frac{\gamma^2(\alpha^2-\alpha)V[\infty]+\alpha
V[1]+(1-\alpha)(1-\gamma^2)V[0]}{1+\gamma^2(\alpha^2-1)}.
\end{eqnarray}
When the expectation value of the kinetic energy in
$\psi^\prime_\alpha$ is included we obtain the following expression
for the energy gap as a function of $\lambda$ and $\alpha$
\begin{equation}
\frac{\Delta[\alpha,\lambda]}{e^2/\epsilon l_0} = \frac{\lambda(1-\gamma^2) +
\gamma^2(\alpha^2-\alpha)V[\infty]+\alpha
V[1]+(1-\alpha)(1-\gamma^2)V[0]}{1+\gamma^2(\alpha^2-1)} - E_0,
\end{equation}
where $E_0 \equiv \langle\psi|V|\psi\rangle$ is the ground state
energy in units of $e^2/\epsilon l_0$.  It is therefore only necessary
to determine five expectation values, $E_0$, $V[0]$, $V[1]$,
$V[\infty]$ and $\gamma$, by variational Monte Carlo.  Once these
quantities have been calculated it is straightforward to minimize
$\Delta[\alpha,\lambda]$ with respect to $\alpha$ and determine the
energy gap as a function of $\lambda$.  In addition, once the value of
$\alpha$ which minimizes the energy of $\psi^\prime_\alpha$ for a
given $\lambda$ is known it is possible to calculate various
expectation values for that value of $\lambda$.

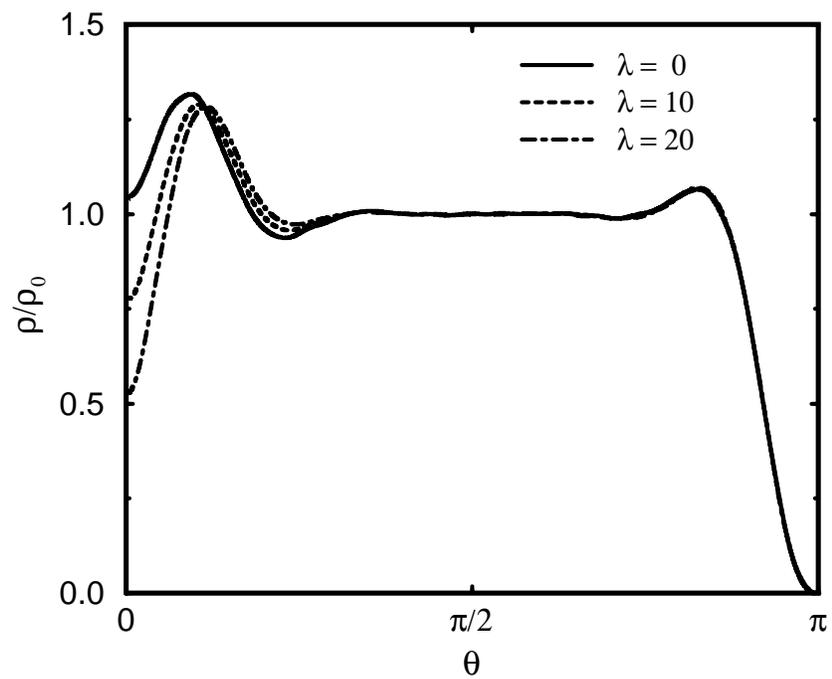
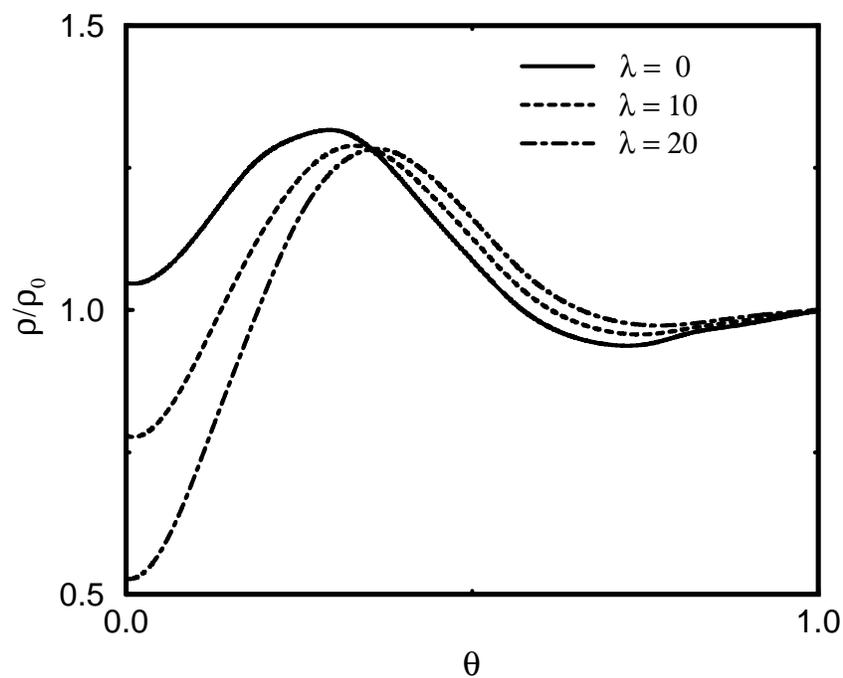
\begin{figure}
\caption{
(Top) Density profile of the excited state wave function on a sphere
with a quasielectron at the top of the sphere $(\theta = 0)$ and a
quasihole on the bottom of the sphere $(\theta = \pi)$ for different
values of the Landau level mixing parameter $\lambda$, and (Bottom)
blow up of the quasielectron density.  As $\lambda$ increases the
quasielectron charge becomes less localized leading to a reduction of
the energy gap.  The results shown are for 30 electrons and $\beta =
0$.}
\label{density}
\end{figure}

Figure \ref{density} shows the density profile of the excited state at
$\nu=1/3$ as a function of $\theta$ for different values of $\lambda$.
The results are shown for $N=30$ and $\beta = 0$.  Note that as
$\lambda$ increases only the quasielectron is affected by the partial
Landau level projection.  This is because Jain's quasihole wave
function is equivalent to Laughlin's and is therefore entirely in the
lowest Landau level.  For $\lambda = 0$ the quasielectron is fully
projected onto the lowest Landau level and its density profile is
nearly indistinguishable from that obtained by Morf and Halperin for
Laughlin's quasielectron trial state
\cite{morf}.  As $\lambda$ increases the quasielectron charge becomes
less localized.  This delocalization occurs because as the first
excited Landau level is mixed into the wave function the quasielectron
has more degrees of freedom allowing it to `spread out' and lower its
Coulomb energy at the price of some kinetic energy.  This
delocalization of the quasielectron charge with increasing $\lambda$
is the physical origin of the reduction of the energy gap by Landau
level mixing.

It is important to point out that the effect of Landau level mixing on
the bulk of the wave function is not included in our calculations and
we do not expect the energy of either $\psi$ or $\psi^\prime_\alpha$
to be valid for finite values of $\lambda$.  However, we do expect the
{\it difference} in these energies to give a reasonable value for the
energy gap, because in this difference the energy associated with the
bulk of the wave function cancels leaving only the energy associated
with the local excitations.  Note that in our calculation the
reduction of the energy gap comes entirely from the quasielectron,
because the quasihole is unaffected by the partial Landau level
projection operator.  This is consistent with Yoshioka's exact
diagonalization calculations which showed that the exciation energy of
the quasielectron was significantly more sensitive to Landau level
mixing than that of the quasihole \cite{yoshioka}.

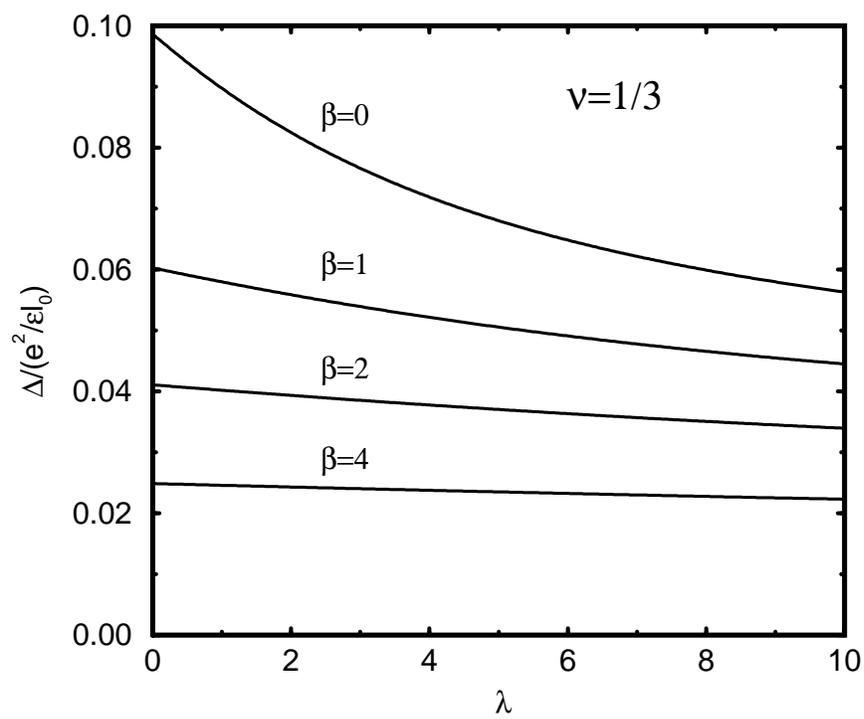
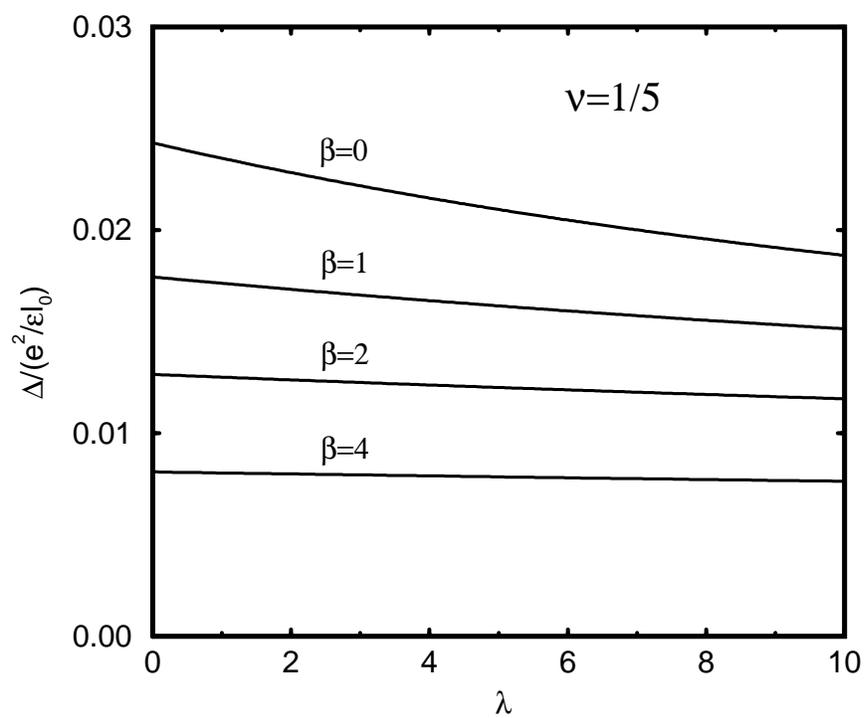
\begin{figure}
\caption{Energy Gap for creating a well separated
quasielectron---quasihole pair at $\nu=1/3$ and $\nu=1/5$ as a
function of the Landau level mixing parameter $\lambda = (e^2/\epsilon
l_0)/\hbar\omega_c$ for various values of the thickness parameter
$\beta$ (defined in the text).  The results shown are for 30
electrons.}
\label{gapvsl}
\end{figure}

Figure \ref{gapvsl} shows the energy gap as a function of $\lambda$
for various values of the thickness parameter $\beta$.  The results
are for a system with 30 electrons.  Both increasing $\beta$ and
increasing $\lambda$ have the effect of reducing the energy gap,
consistent with previous calculations \cite{yoshioka,zhang}.  The key
result of our calculation is the observation that, as $\beta$
increases, the effect of Landau level mixing on the energy gap becomes
weaker.  This weakening of the Landau level mixing effect can be
understood as follows. The thickness correction softens the
short-range part of the Coulomb interaction, which in turn reduces the
interaction energy the quasielectron stands to gain by delocalizing.
It follows that as $\beta$ increases the quasielectron charge
delocalizes less for a given value of $\lambda$, and the energy gain
also decreases.  Note that for the experimentally relevant value
$\beta \simeq 1$ the energy gap is not much different for $\lambda=1$
and $\lambda=5$.  For $\lambda = 1$ the energy gap is roughly $0.06
e^2/\epsilon l_0$, consistent with the experimentally measured value
of $\Delta_e \simeq 0.05 e^2/\epsilon l_0$, but for $\lambda = 5$ the
theoretical energy gap is much larger than the experimentally observed
$\Delta_h \simeq 0.023 e^2/\epsilon l_0$.  We must therefore conclude
that Landau level mixing alone cannot account for the reduced energy
gap observed in $p$-type GaAs quantum wells.

To conclude, we have modified Jain's composite fermion wave functions
by applying a partial Landau level projection operator.  These new
wave functions have been used to calculate the effect of Landau level
mixing on the energy gap in the fractional QHE.  The main result is
the observation that as the thickness parameter $\beta$ is increased
the effect of Landau level mixing on the energy gap is suppressed. In
particular, for the experimentally relevant value of $\beta \simeq 1$
Landau level mixing has almost no effect on the energy gap.  Thus,
even though our results are in good agreement with the experimentally
measured energy gap in $n$-type GaAs systems at $\nu=1/3$, we conclude
that Landau level mixing alone cannot account for the factor of two
smaller energy gaps observed in $p$-type GaAs quantum wells.

This work was supported by the National High Magnetic Field Laboratory
at Florida State University.

\begin{references}

\bibitem{qhe} For a review see {\it The Quantum Hall Effect},
edited by R.E. Prange and S.M. Girvin (Springer-Verlag, New York,
1990).

\bibitem{jain}
J.K. Jain, Phys. Rev. Lett. {\bf 63} (1989) 199; {\it Phys. Rev.  B}
{\bf 40}, 8079 (1989); {\it ibid.} {\bf 41}, 7653 (1990).

\bibitem{trivedi}
N. Trivedi and J. Jain, Mod. Phys. Lett. B {\bf 5} 503 (1991).

\bibitem{neb}
N.E. Bonesteel, Phys. Rev. B {\bf 51}, 9917 (1995).


\bibitem{manoharan}
H.C. Manoharan, M. Shayegan, and S.J. Klepper, Phys. Rev. Lett. {\bf
73}, 3270 (1994).

\bibitem{du}
See, for example, R.R. Du {\it et al.}, {\it Phys.  Rev. Lett.} {\bf
70}, 2944 (1992).

\bibitem{willett}
R.L. Willett, H.L. Stormer, D.C. Tsui, A.C. Gossard, and J.H. English,
Phys. Rev. B {\bf 37}, 8476 (1988).

\bibitem{yoshioka}
D. Yoshioka, J. Phys. Soc. Jpn. {\bf 55}, 885 (1986).

\bibitem{haldane}
F.D.M. Haldane, Phys. Rev. Lett. {\bf 51}, 605 (1983).

\bibitem{dev}
G. Dev and J.K. Jain, Phys. Rev. Lett. {\bf 69}, 2843 (1992);
R.K. Kamilla, Z.G. Wu, and J.K. Jain, Preprint.

\bibitem{ceperely}
D. Ceperley, G.V. Chester, and M.H. Kalos, Phys. Rev. B {\bf 16}, 3081
(1977).

\bibitem{zhang}
F.C. Zhang and S. Das Sarma, Phys. Rev. B {\bf 33}, 2903 (1986).

\bibitem{morf} R. Morf and B.I. Halperin, Phys. Rev. B {\bf 37}, 8476 (1988).

\end{references}
\end{document}